\documentstyle[12pt,psfig]{article}
\newcommand{\EQ}{\begin{equation}}
\newcommand{\EN}{\end{equation}}
\newcommand{\bea}{\begin{eqnarray}}
\newcommand{\eea}{\end{eqnarray}}

\newcommand{\th}{\theta}

\begin{document}
\topmargin 0pt
\oddsidemargin 5mm
\renewcommand{\thefootnote}{\arabic{footnote}}
\newpage
\setcounter{page}{0}
\begin{titlepage}
\hspace{11.8cm}SISSA-24/2001/FM
\vspace{1cm}
\vspace{0.5cm}
\begin{center}
{\large {\bf Sine-Gordon description of the scaling three-state Potts 
antiferromagnet on the square lattice}}\\
\vspace{1.8cm}
{\large Gesualdo Delfino}\footnote{E-mail: delfino@sissa.it} \\ \vspace{0.5cm}
{\em International School for Advanced Studies, via Beirut 4, 34014 Trieste, 
Italy \\
and Istituto Nazionale di Fisica Nucleare, Sezione di Trieste}
\end{center}
\vspace{1.2cm}

\renewcommand{\thefootnote}{\arabic{footnote}}
\setcounter{footnote}{0}

\begin{abstract}
\noindent
The scaling limit as $T\rightarrow 0$ of the antiferromagnetic three-state 
Potts model on the square lattice is described by the sine-Gordon quantum 
field theory at a specific value of the coupling. We show that the 
correspondence follows unambigously from an analysis of the sine-Gordon 
operator space based on locality, and that the scalar operators carrying 
solitonic charge play an essential role in the description of the lattice 
model. We then evaluate the correlation functions within the form factor 
approach and give a number of universal predictions that can be checked 
in numerical simulations.
\end{abstract}

\vspace{.3cm}

\end{titlepage}
\newpage
The study of antiferromagnetic models is a notoriously difficult problem
of statistical mechanics. The dependence on the lattice structure
produces a variety of behaviours much richer than in the ferromagnetic 
case and forces a case-by-case analysis. Many antiferromagnetic models 
possess a critical point (often at zero temperature), so that their continuum
limit can also be investigated through field theoretical methods. Among
them, the square lattice three-state Potts model has been the 
object of both numerical and theoretical studies. While this model has been
known for longtime to be critical at zero temperature, the issue of the 
approach to criticality presents several subtleties, including the 
identification of the correct scaling variable. Very recently, the authors
of reference \cite{CJS} exploited a mapping onto a height model to 
identify the excitations on the lattice at nonzero temperature and explain
the anomalous corrections to scaling previously observed in Monte Carlo
studies \cite{FS}.

It is the purpose of this note to point out that the {\em scaling limit}
of the square lattice three-state Potts antiferromagnet is {\em exactly}
solvable due to its equivalence to a specific point of the sine-Gordon
model (an integrable quantum field theory), and to derive from this
continuum approach some universal predictions that can be checked through
simulations on the lattice.

As it often happens for two-dimensional systems exhibiting a gaussian critical 
point, a relation of the scaling limit with the sine-Gordon model is expected.
The actual task is that of understanding what the symmetries and the 
operators of the lattice model become in the field theoretic 
language. Here we will show that these identifications follow in a quite 
natural way from the analysis of the sine-Gordon operator space based on 
locality, and that the topologically charged scalar operators play a 
major role in the continuum description of the lattice model. Matrix 
elements for these operators were first computed in \cite{AT}, where they
arise in the description of the scaling Ashkin-Teller model. We will then
illustrate how the correspondence at the operatorial level translates 
in the language of particle excitations, in order to exploit the exact 
$S$-matrix solution of the sine-Gordon model and extract from that the 
correlation functions for the operators of physical interest.
In particular, we will give a number of specific and universal predictions
suitable for a numerical verification on the lattice.

\vspace{.3cm}
The three-state Potts model is defined by the lattice Hamiltonian
\EQ
H=-J\sum_{\langle i,j\rangle}\delta_{s_i,s_j}\,\,,
\label{hamiltonian}
\EN
where the spin variable $s_j$ on the $j$-th lattice site takes the values 
(colours) $0$,$1$,$2$ and the sum is over nearest neighbours. The Hamiltonian
is invariant under permutations of the colours. Here we are interested
in the antiferromagnetic case ($J<0$) on the square lattice.

At zero temperature, the system is in one of the ground states in which
nearest neighbours spins have different values. This problem is equivalent
to the three colouring problem of the square lattice which admits an exact
mapping onto a specific point of the six-vertex model \cite{LW,Baxter}. From 
the exact solution of the latter we know that our zero-temperature system is 
critical and that its long-distance behaviour is described by a free massless 
boson. Relevant operators
that can be identified through the zero-temperature analysis on the lattice
\cite{dNNS,PW,BH,SS} are the staggered magnetisation 
$\Sigma_j=(-1)^{j_1+j_2}\,e^{2i\pi s_j/3}$ with scaling dimension
$1/6$, the uniform magnetisation $\sigma_j=e^{2i\pi s_j/3}$ with scaling
dimension $2/3$, and the staggered polarisation\footnote{The primed sum 
indicates summation over the next nearest neighbours of $j$.} 
${\cal P}_j=(-1)^{j_1+j_2}\sum'_i(2\delta_{s_i,s_j}-1)$ with 
scaling dimension $3/2$. We identify the $j$-th site of the
square lattice through a pair of integers $(j_1,j_2)$, and call even (odd)
sublattice the collection of the sites with $j_1+j_2$ even (odd). 

At non-zero temperature, the model develops a finite correlation length
and its scaling limit is described by the perturbation
of the gaussian fixed point through the continuum version of the thermal
operator ${\cal E}_j=\sum_i\delta_{s_i,s_j}$. This perturbed theory
is the sine-Gordon model defined by the euclidean action\footnote{Throughout
this note we use the standard notation $\beta$ for the sine-Gordon 
dimensionless coupling. No confusion with the inverse temperature of the 
lattice model should be made.}
\EQ
{\cal A}=\int d^2x\,\left(\frac12\,\partial_\alpha\varphi\partial^\alpha
\varphi-\mu\cos\beta\varphi\right)\,\,,
\label{action}
\EN
for some value of the coupling $\beta$ to be determined. The theory 
(\ref{action}) 
describes the scaling limit of several lattice models whose critical 
point corresponds to a conformal field theory with central charge $C=1$
(see e.g. Refs.\,\cite{AT,q4} for other examples discussed in the framework
of this note). In order to proceed with the description of the 
antiferromagnetic Potts model we need to recall few points about the 
operator content of the sine-Gordon model.

\vspace{.3cm}
At criticality ($\mu=0$), the boson field can be decomposed into its
holomorphic and antiholomorphic parts as $\varphi(x)=
\phi(z)+\bar{\phi}(\bar{z})$, where we introduced the complex coordinates
$z=x_1+ix_2$ and $\bar{z}=x_1-ix_2$. The scaling operators of the theory 
are the vertex operators
\EQ
V_{p,\bar{p}}(x)=e^{i[p\phi(z)+\bar{p}\bar{\phi}(\bar{z})]}\,\,,
\label{vertex}
\EN
with conformal dimensions $(\Delta,\bar{\Delta})=(p^2/{8\pi},\bar{p}^2/{8\pi})$
and spin $s=\Delta-\bar{\Delta}$. They satisfy the gaussian operator
product expansion
\EQ
V_{p_1,\bar{p}_1}(x)V_{p_2,\bar{p}_2}(0)=
z^{p_1p_2/{4\pi}}\bar{z}^{\bar{p}_1\bar{p}_2/{4\pi}}\,
V_{p_1+p_2,\bar{p}_1+\bar{p}_2}(0)+\ldots\,\,\,.
\label{ope}
\EN
We see from this relation that taking $V_{p_1,\bar{p}_1}(x)$ around 
$V_{p_2,\bar{p}_2}(0)$ by sending $z\rightarrow ze^{2i\pi}$ and 
$\bar{z}\rightarrow \bar{z}e^{-2i\pi}$ produces a phase factor
$e^{2i\pi\gamma_{1,2}}$, where
\EQ
\gamma_{1,2}=\frac{1}{4\pi}\,(p_1p_2-\bar{p}_1\bar{p}_2)
\label{gamma}
\EN
is called index of mutual locality. If $\gamma_{1,2}$ is an integer the 
correlator $\langle V_{p_1,\bar{p}_1}(x)V_{p_2,\bar{p}_2}(0)\rangle$ is
single valued and the two operators are said to be mutually local. 
Since $\gamma_{1,1}=2s$, the operators
which are local with respect to themselves (the only ones we are interested
in here) must have integer or half integer spin.

In the off-critical theory (\ref{action}), the operators wich are local
with respect to the perturbing operator
$\cos\beta\varphi\sim V_{\beta,\beta}+V_{-\beta,-\beta}$ form a `local
sector' into which all the operators of interest for the description of the
lattice model are expected to fall. This locality requirement
selects the operators $V_{p,\bar{p}}$ with $p-\bar{p}=4\pi m/\beta$, 
$m$ integer, namely
\EQ
V_p(x)\equiv V_{p,p}(x)=e^{ip\varphi(x)}\,\,,
\label{v}
\EN
and 
\EQ
U_{n,m}(x)\equiv V_{\frac{n\beta}{2m}+\frac{2\pi}{\beta}m,
                    \frac{n\beta}{2m}-\frac{2\pi}{\beta}m}(x)
=e^{i\left[\frac{n\beta}{2m}\varphi(x)+\frac{2\pi}{\beta}m
\tilde{\varphi}(x)\right]}\,,\hspace{.4cm}n=2s=0,\pm 1,\ldots,
\hspace{.3cm}m=\pm 1,\ldots
\label{u}
\EN
Here we introduced the `dual' boson field $\tilde{\varphi}$ which is 
$\phi(z)-\bar{\phi}(\bar{z})$ at criticality and satisfy the relation 
\EQ
i\frac{\partial\tilde{\varphi}}{\partial x_\alpha}=\varepsilon_{\alpha\beta}
\frac{\partial\varphi}{\partial x_\beta}\,\,.
\label{dual}
\EN

The operators $V_p$ and $U_{0,m}$ are scalars ($s=0$) and have scaling
dimensions $X_p=p^2/4\pi$ and $X_{0,m}=\pi m^2/\beta^2$, respectively 
($X=\Delta+\bar{\Delta}$). The action (\ref{action}) describes a relevant
perturbation of the gaussian fixed point for $\beta^2<8\pi$; in this range
the only operators $U_{0,m}$ which are relevant ($X_{0,m}<2$) are those 
with $|m|\leq 3$.

The lowest operators with $|s|=1/2$, i.e. $\Psi=U_{\pm 1,1}$
and $\Psi^*=U_{\pm 1,-1}$, are complex conjugated two-component spinors with 
conformal dimensions $\Delta_{n,m}$ given by
\bea
&& \Delta_{\pm 1,1}=\Delta_{\pm 1,-1}=\frac{1}{8}\left(\frac{\beta^2}{4\pi}
\pm 2+\frac{4\pi}{\beta^2}\right)\,,\label{w1}\\
&& \bar{\Delta}_{\pm 1,1}=\bar{\Delta}_{\pm 1,-1}=
\frac{1}{8}\left(\frac{\beta^2}{4\pi}\mp 2+\frac{4\pi}{\beta^2}\right)\,.
\label{w2}
\eea
It is known since Coleman \cite{Coleman} that the sine-Gordon model is 
equivalent to the theory of a Dirac fermion with four fermion interaction,
the Thirring model. The expression for $\Psi$ concides with the
bosonisation formula for the Thirring fermion originally derived by Mandelstam 
\cite{Mandelstam}. The value 
$\beta^2=4\pi$ for which the dimensions (\ref{w1}), (\ref{w2}) take the free 
fermionic values corresponds to the free point of the Thirring model. 

It follows from these considerations that the integer $m$ in (\ref{u})
is a fermionic charge. Since the Thirring fermions correspond to the 
solitons interpolating between adjacent vacua of the periodic bosonic
potential, we will also call $m$ topologic charge. Hence, we can make a 
distinction between neutral scalar operators $V_p(x)$ with a non-zero
vacuum expectation value, and charged scalar operators $U_{0,m}(x)$.

\vspace{.3cm}
Going back to the Potts model it is not difficult to identify the 
continuum limit of the lattice operators among the sine-Gordon operators. 
Both the staggered and uniform magnetisation are charged with respect to
colour permutations and must be found among the $U_{0,m}$. Comparison
with the known scaling dimensions gives $\Sigma(x)\sim U_{0,1}(x)$ 
and $\sigma(x)\sim U_{0,-2}(x)$, and selects
\EQ
\beta=\sqrt{6\pi}\,\,,
\EN
as the value for which the sine-Gordon model (\ref{action}) describes
the scaling limit of the three-state Potts antiferromagnet on the square
lattice. This immediately fixes the scaling dimension of the thermal
operator ${\cal E}(x)\sim\cos\beta\varphi(x)$ to be $3/2$, in agreement with 
the result obtained in \cite{CJS} by studing the vortex excitations on the 
lattice. 

On the basis of these identifications we can see how the symmetries of the 
lattice model translate in the sine-Gordon language. The group $S_3$ of
colour permutations can be decomposed into the $Z_3$ transformations
associated to cyclic permutations plus the complex conjugation of 
$e^{2i\pi s_j/3}$. The latter operation simply corresponds to the 
complex conjugation of the sine-Gordon exponentials, while the 
elementary $Z_3$ transformation maps into the shift 
$\tilde{\varphi}\rightarrow\tilde{\varphi}+2\pi/\beta$. The $Z_3$ charge
coincides with the topologic charge $m$ (mod $3$).

The lattice 
operators are also characterised by their parity under the transformation 
which exchanges the even and odd sublattices: ${\cal E}_j$ and $\sigma_j$
are even, while $\Sigma_j$ and ${\cal P}_j$ are odd. It appears
that in the continuum limit this parity property corresponds to $(-1)^m$. 
The staggered polarisation
${\cal P}_j$ is invariant under colour permutations and must correspond
to the operator $U_{0,3}+U_{0,-3}$, which has indeed the expected 
scaling dimension $3/2$. We summarise the situation in Table\,1.

\begin{center}

\vspace{1cm}
\begin{tabular}{|c|c|c|c|c|}\hline
$\Phi$ & Lattice & Continuum & $X_\Phi$ & Topologic \\
& definition & limit & & charge \\ \hline
${\cal E}$ & $\sum_i\delta_{s_i,s_j}$ & $\cos\sqrt{6\pi}\varphi$ & $3/2$ & 
$0$ \\
$\Sigma$ & $(-1)^{j_1+j_2}\,e^{2i\pi s_j/3}$ & $e^{i\sqrt{2\pi/3}\,
\tilde{\varphi}}$ & $1/6$ & $1$ \\
$\sigma$ & $e^{2i\pi s_j/3}$ & $e^{-2i\sqrt{2\pi/3}\,\tilde{\varphi}}$ & 
$2/3$ & $-2$ \\
${\cal P}$ & $(-1)^{j_1+j_2}\sum'_i(2\delta_{s_i,s_j}-1)$ & 
$\cos\sqrt{6\pi}\,\tilde{\varphi}$ & $3/2$ & $\pm 3$ \\ \hline
\end{tabular}
\end{center}
\begin{center}
{\bf Table 1}
\end{center}

\vspace{.5cm}
The equivalence with a particular case of the sine-Gordon model ensures that 
the scaling limit of the lattice model is an {\em integrable} quantum field 
theory whose associated scattering theory is known exactly \cite{ZZ}. 
The elementary excitations are a pair of conjugated particles $A_+$ and
$A_-$ (the sine-Gordon soliton and antisoliton). The
fact that $\beta^2=6\pi$ falls in the repulsive sine-Gordon regime 
$\beta^2>4\pi$ 
ensures that no other particles are present in the spectrum. It can be 
interesting to mention that the $Z_3$-preserving fusion $A_+A_+\rightarrow
A_-$ is forbidden here because it violates the sublattice parity
$(-1)^m$; it is instead characteristic of the scattering theory of the 
ferromagnetic case \cite{ferro1,ferro2} in which the lattice plays no role. 

Due to the factorisation of multiparticle scattering in the 
integrable quantum field theories, the scattering theory is completely
determined by the two-particle $S$-matrix defined by the relation\footnote{
The rapidity $\th$ parameterises the on-shell energy and momentum of a 
particle as $(p^0,p^1)=(m\cosh\th,m\sinh\th)$.} 
\EQ
A_a(\th_1)A_b(\th_2)=\sum_{c,d=\pm}S_{ab}^{cd}(\th_1-\th_2)
A_d(\th_2)A_c(\th_1)\,, \hspace{1cm}a,b=\pm\,.
\label{FZ}
\EN                                                               
The non-zero scattering amplitudes are given by\footnote{For reasons that 
will become clear later, it is useful to give the results referring to 
generic values of $\beta$ in the sine-Gordon model, being understood that 
the scaling limit we are dealing with corresponds to $\beta=\sqrt{6\pi}$.} 
\cite{ZZ}
\bea
&& S_{++}^{++}(\th)=S_{--}^{--}(\th)=S_0(\th)\,,\\
&& S_{+-}^{+-}(\th)=S_{-+}^{-+}(\th)=-\frac{\sinh\frac{\pi\th}{\xi}}
               {\sinh\frac{\pi}{\xi}(\th-i\pi)}\,S_0(\th)\,,\\
&& S_{+-}^{-+}(\th)=S_{-+}^{+-}(\th)=-\frac{\sinh\frac{i\pi^2}{\xi}}
               {\sinh\frac{\pi}{\xi}(\th-i\pi)}\,S_0(\th)\,,
\eea
with
\EQ
S_0(\th)=-\exp\left\{-i\int_0^\infty\frac{dx}{x}\frac{\sinh\frac{x}{2}\left(1-
\frac{\xi}{\pi}\right)}{\sinh\frac{x\xi}{2\pi}\cosh\frac{x}{2}}
\sin\frac{\th x}{\pi}\right\}\,\,,
\EN
\EQ
\xi=\frac{\pi\beta^2}{8\pi-\beta^2}\,\,.
\EN 
From the $S$-matrix one can compute the form factors\footnote{$|0\rangle$
is the vacuum state.}
\EQ
F^\Phi_{a_1,\ldots,a_n}(\th_1,\ldots,\th_n)=\langle 0|\Phi(0)|A_{a_1}(\th_1),
\ldots,A_{a_n}(\th_n)\rangle\,,
\hspace{1cm}a_i=\pm
\label{ff}
\EN
which in turn determine the spectral decomposition of the correlation 
functions. The form factors satisfy the equations 
\cite{Karowski,Smirnov,YZ}
\bea
&& F^\Phi_{a_1,\ldots,a_i,a_{i+1},\ldots,a_n}(\th_1,\ldots,\th_i,\th_{i+1},
\ldots,\th_n)=\nonumber \\
&& \sum_{b_i,b_{i+1}=\pm}S_{a_i,a_{i+1}}^{b_i,b_{i+1}}
(\th_i-\th_{i+1})F^\Phi_{a_1,\ldots,b_{i+1},b_i,\ldots,a_n}
(\th_1,\ldots,\th_{i+1},\th_i,\ldots,\th_n)\,\,,
\label{ff1}
\eea
\EQ
F^\Phi_{a_1,\ldots,a_n}(\th_1+2i\pi,\th_2,\ldots,\th_n)=
e^{2i\pi\gamma_{\Phi,a_1}}F^\Phi_{a_2,\ldots,a_n,a_1}(\th_2,\ldots,\th_n,
\th_1)\,\,,
\label{ff2}
\EN
where $\gamma_{\Phi,a}$ in the last equation is the index of mutual locality
between the scalar operator $\Phi(x)$ and the soliton (for $a=+$) or the 
antisoliton (for $a=-$). Since these particles are created by the operators
$U_{0,\pm 1}(x)$, equation (\ref{gamma}) gives for the scalar operators
\EQ
\gamma_{V_p,\pm}=\gamma_{V_p;U_{0,\pm 1}}=\pm\frac{p}{\beta}\,\,,
\label{g1}
\EN
\EQ
\gamma_{U_{0,m};\pm}=\gamma_{U_{0,m};U_{0,\pm 1}}=0\,\,.
\label{g2}
\EN

Equations (\ref{ff1}) and (\ref{ff2}) can be used to determine the `lowest'
form factors for the operator $\Phi$, i.e. the non-vanishing matrix 
elements (\ref{ff}) with the smallest $n$ ($n>0)$ which fix the initial 
conditions of the form factor bootstrap\footnote{The higher form 
factors contain additional pairs $A_+(\th)A_-(\th')$ in the asymptotic 
state and are related to the lowest ones by a recursive equation associated
to particle-antiparticle annihilation \cite{Smirnov}.}.
The lowest form factors for the operators of interest in this note are
\EQ
F^{\cos\beta\varphi}_{\pm\mp}(\th_1,\th_2)=c_0\,
\frac{\cosh\frac{\th_{12}}{2}}
{\sinh\frac{\pi}{2\xi}(\th_{12}-i\pi)}\,F_0(\th_{12})\,\,,
\label{cos}
\EN
\EQ
F^{U_{0,m}}_{-sg(m),\ldots,-sg(m)}(\th_1,\ldots,\th_{|m|})=
c_{|m|}\prod_{i<j}F_0(\th_{ij})\,\,,
\label{charged}
\EN
where $\th_{ij}\equiv\th_i-\th_j$, $sg(m)$ denotes the sign of $m$, and 
the $c_m$ are normalisation constants. The function
\EQ
F_0(\th)=-i\sinh\frac{\th}{2}\,
\exp\left\{\int_0^\infty\frac{dx}{x}\,\frac{\sinh\left[\frac{x}{2}\left(1-
\frac{\xi}{\pi}\right)\right]}{\sinh\frac{x\xi}{2\pi}\,\cosh\frac{x}{2}}\,
\frac{\sin^2\frac{(i\pi-\th)x}{2\pi}}{\sinh x}\right\}\,\,
\EN
satisfies the equations 
\bea
&& F_0(\th)=S_0(\th)F_0(-\th)\,\,,\\
&& F_0(\th+2i\pi)=F_0(-\th)\,\,.
\eea

The lowest form factors determine the first term in the large distance 
expansion of the correlation functions. Using the subscript $c$ to denote
the connected correlators we have
\bea
\langle V_p(x)V_p(0)\rangle_c &=& \int\frac{d\th_1}{2\pi}\frac{d\th_2}{2\pi}\,
|F^{V_p}_{+,-}(\th_1,\th_2)|^2\,e^{-M|x|(\cosh\th_1+\cosh\th_2)} \nonumber \\ 
&+& O(e^{-4M|x|})\,\,,\label{corrv}\\
\langle U_{0,m}(x)U_{0,-m}(0)\rangle &=& \frac{1}{m!}
\int\frac{d\th_1}{2\pi}\cdots
\frac{d\th_m}{2\pi}\,|F^{U_{0,m}}_{-,\ldots,-}(\th_1,\ldots,\th_m)|^2\,
e^{-M|x|\sum_{i=1}^m\cosh\th_i} \nonumber \\
&+& O(e^{-(m+2)M|x|})\,\,,
\label{corru}
\eea
where $M$ is the mass of the soliton and we took $m>0$.

\vspace{.3cm}
There are a number of simple and distinguished universal predictions of the 
sine-Gordon description that can be tested in 
numerical simulations of the square lattice three-state Potts antiferromagnet. 
The `exponential' and `second moment' correlation 
lengths $\xi_\Phi$ and $\xi^{2nd}_\Phi$ associated to an operator $\Phi(x)$ are
defined as
\EQ
\langle\Phi(x)\Phi^*(0)\rangle_c\sim\exp(-|x|/\xi_\Phi)\,,\hspace{1cm}
|x|\rightarrow\infty\,,
\EN
\EQ
\xi^{2nd}_\Phi=\left(\frac14\,\frac{\int d^2x\,|x|^2\,
\langle\Phi(x)\Phi^*(0)\rangle_c}{\int d^2x\,\langle\Phi(x)\Phi^*(0)\rangle_c}
\right)^{1/2}\,.
\EN
The leading behaviour of these quantities nearby the critical point is
\bea
\xi_\Phi\simeq f_\Phi\,t^{-\nu}\,\,,\\
\xi^{2nd}_\Phi\simeq f^{2nd}_\Phi\,t^{-\nu}\,\,,
\eea
where $\nu=1/(2-X_{\cal E})=2$, and 
$t$ measures the deviation from criticality. It appears from 
numerical simulations \cite{FS} and has been discussed in \cite{CJS} that
for this zero-temperature critical point the scaling variable is $t=e^{J/kT}$. 
Our operator identifications and equations (\ref{corrv}), (\ref{corru}) then 
give $\xi_\Sigma=1/M$ and
\bea
f_\sigma/f_\Sigma &=& \frac12\,\,,\label{rapp1}\\
f_{\cal E}/f_\Sigma &=& \frac12\,\,,\label{rapp2}\\
f_{\cal P}/f_\Sigma &=& \frac13\,\,,\label{rapp3}\\
f^{2nd}_\Sigma/f_\Sigma &\approx & 1\,\,,\label{rapp4}\\
f^{2nd}_\sigma/f_\sigma &\approx & 0.439\,\,.\label{rapp5}
\eea
The last two ratios are evaluated in the lowest form factor approximation
which is known to give estremely accurate results for integrated 
correlators\footnote{See e.g. \cite{AT,q4} for similar computations in other
sine-Gordon related statistical models.}. We estimate that our error on 
these two quantities does not exceed 1\%. 

The universal ratios (\ref{rapp1})--(\ref{rapp5}) 
have not yet been measured in simulations. Lattice estimates of these 
quantities would provide the direct confirmation that the universality 
class of this antiferromagnet is described by the sine-Gordon field theory
with $\beta=\sqrt{6\pi}$ and the operator identifications of Table\,1.
In this respect, it is important to mention that the study of the scaling
limit of the model on the lattice is complicated in practice by strong 
corrections to
scaling \cite{CJS,FS}. Cardy {\em et al.} argued in \cite{CJS} that such 
corrections are originated by a strictly marginal operator which couples to 
the temperature giving an effective thermal dependence to the stiffness $K$ of 
the bosonic field, which is related to the sine-Gordon coupling as 
$\beta=6\sqrt{K}$. Their estimate of this (nonuniversal) dependence for 
$K_{eff}(t)$ corresponds to $\beta_{eff}^2(t)=6\pi-20.9(4)t-37.8(2)t^2+\ldots$.
Hence, according to this analysis, the leading corrections to scaling
can be fitted by letting the critical exponents and the critical amplitudes 
vary with $\beta_{eff}$. Concerning the lattice verification of our 
predictions,
the ratios (\ref{rapp1})--(\ref{rapp4}) are not affected by this effect 
since their values in the sine-Gordon model are determined by symmetry 
properties of the operators which do not depend on the 
coupling\footnote{At least in the range $\beta^2\geq 4\pi$ in which there 
are no bound states.}. The ratio $(f^{2nd}_\sigma/f_\sigma)_{eff}$ is 
instead an increasing function of $\beta_{eff}$, so that the asymptotic value
(\ref{rapp5}) should be approached from below as $t\rightarrow 0$. For 
example, we find $0.38$ when computing this ratio at 
$\beta_{eff}^2(t=0.1)\sim 6\pi-2.47$.

\vspace{.7cm}
{\bf Acknowledgments:} I thank John Cardy for helpful discussions and remarks.

\vspace{.6cm}
{\bf Note added:} When this work had been completed reference \cite{LZ}
appeared which is devoted to the form factors of the sine-Gordon
topologically charged operators and whose main purpose is the determination
of their ``conformal'' normalisation.


\begin{thebibliography}{99}

\bibitem{CJS} J.L. Cardy, J.L. Jacobsen and A.D. Sokal, Unusual
corrections to scaling in the 3-state Potts antiferromagnet on a square
lattice, cond-mat/0101197.
\bibitem{FS} S.J. Ferreira and A.D. Sokal, {\em J. Stat. Phys.} {\bf 96}
(1999) 461.
\bibitem{AT} G. Delfino, {\em Phys. Lett.} {\bf B 450} (1999) 196.
\bibitem{LW} E.H. Lieb and F.Y. Wu, in {\em Phase Transitions and 
Critical Phenomena}, edited by C. Domb and M.S. Green (Academic, New
York, 1972), Vol. 1.
\bibitem{Baxter} R.J. Baxter, Exactly solved models of statistical
mechanics, Academic Press, London, 1982.
\bibitem{dNNS} M. den Nijs, M.P. Nightingale and M. Schick, {\em Phys.
Rev.} {\bf B 26} (1982) 2490.
\bibitem{PW} H. Park and M. Widom, {\em Phys. Rev. Lett.} {\bf 63} (1989)
1193.
\bibitem{BH} J.K. Burton Jr. and C.L. Henley, {\em J. Phys.} {\bf A 30}
(1997) 8385.
\bibitem{SS} J. Salas and A.D. Sokal, {\em J. Stat. Phys.} {\bf 92} (1998)
729.
\bibitem{q4} G. Delfino and J.L. Cardy, {\em Phys. Lett.} {\bf B 483}
(2000) 303.
\bibitem{Coleman} S. Coleman, {\em Phys. Rev.} {\bf D 11} (1975) 2088.
\bibitem{Mandelstam} S. Mandelstam, {\em Phys. Rev.} {\bf D 11} (1975)
3026.
\bibitem{ZZ} A.B. Zamolodchikov and Al.B. Zamolodchikov, {\em Ann. Phys.} 
{\bf 120} (1979), 253.
\bibitem{ferro1} R. Koberle and J.A. Swieca, {\em Phys. Lett.} {\bf B 86}
(1979) 209.
\bibitem{ferro2} A.B. Zamolodchikov, {\em Int. J. Mod. Phys.} {\bf A 3}
(1988) 743.
\bibitem{Karowski} M. Karowski and P. Weisz, {\em Nucl. Phys.} {\bf B 139}
(1978), 455.
\bibitem{Smirnov} F.A. Smirnov, Form Factors in Completely Integrable
Models of Quantum Field Theories (World Scientific) 1992.
\bibitem{YZ} V.P. Yurov and Al.B. Zamolodchikov, {\em Int. J. Mod. Phys.}
{\bf A 6} (1991) 3419.
\bibitem{LZ} S. Lukyanov and A. Zamolodchikov, Form factors of
soliton creating operators in the sine-Gordon model, hep-th/0102079.

\end{thebibliography}
\end{document}